From Biometrics to Environmental Control: AI-Enhanced Digital Twins for Personalized Health Interventions in Healing Landscapes

Yiping Meng[1], Yiming Sun[2]

[1]School of Computing, Engineering and Digital Technologies, Teesside University, Middlesbrough, Tees Valley TS1 3BX, UK
[2]School of Electrical and Electronic Engineering, University of Sheffield, Western Bank, Sheffield S10 2TN, UK

Abstract:

The dynamic nature of human health and comfort calls for adaptive systems that respond to individual physiological needs in real time. This paper presents an AI-enhanced digital twin framework that integrates biometric signals, specifically electrocardiogram (ECG) data, with environmental parameters such as temperature, humidity, and ventilation. Leveraging IoT-enabled sensors and biometric monitoring devices, the system continuously acquires, synchronises, and preprocesses multimodal data streams to construct a responsive virtual replica of the physical environment. To validate this framework, a detailed case study is conducted using the MIT-BIH noise stress test dataset. ECG signals are filtered and segmented using dynamic sliding windows, followed by extracting heart rate variability (HRV) features such as SDNN, BPM, QTc, and LF/HF ratio. Relative deviation metrics are computed against clean baselines to quantify stress responses. A random forest classifier is trained to predict stress levels across five categories, and Shapley Additive exPlanations (SHAP) is used to interpret model behaviour and identify key contributing features. These predictions are mapped to a structured set of environmental interventions using a Five-Level Stress–Intervention Mapping, which activates multi-scale responses across personal, room, building, and landscape levels. This integration of physiological insight, explainable AI, and adaptive control establishes a new paradigm for health-responsive built environments. It lays the foundation for the future development of intelligent, personalised healing spaces.

1. Introduction

In an era of rapid urbanization, the challenges of dense living, noise pollution, and environmental stressors have underscored the critical need for spaces that actively promote public health and well-being. Healing landscapes, thoughtfully designed environments that leverage natural elements to deliver restorative benefits, are increasingly vital in urban planning, healthcare facilities, and workplaces. These landscapes are not only visually appealing but also purposefully engineered to alleviate the psychological and physiological burdens of urban life, fostering solace, recovery, and resilience [1], [2].

Research in environmental psychology highlights the profound influence of spatial quality on mental and physical health. The biophilia hypothesis, which posits an innate human connection to nature, provides a theoretical foundation for incorporating natural elements into design [3]. Empirical evidence supports these claims, demonstrating that exposure to natural features, such as greenery, water, or natural light, can significantly reduce stress, enhance cognitive performance, and accelerate physical recovery. For example, office environments integrating biophilic elements, such as indoor plants and water features, have been shown to lower stress levels and increase job satisfaction, with studies reporting up to a 15% improvement in employee well-being [4].

However, traditional healing landscapes often suffer from a critical limitation which is the static nature. Designed and implemented at a fixed point in time, these environments lack the flexibility to adapt to users' evolving needs or respond to dynamic external conditions, such as fluctuating temperatures, lighting, or noise levels. This rigidity is particularly challenging in urban settings, where spatial constraints and rapidly changing environmental factors demand innovative, adaptable solutions to maintain optimal conditions for health and well-being [5] .

Emerging technologies, notably digital twins and artificial intelligence (AI), offer transformative potential to address these shortcomings. Digital twin, a dynamic virtual replica of a physical system, enables real-time monitoring, simulation, and optimisation of environmental conditions [6]. When integrated with AI and Internet of Things (IoT) sensors, digital twins can analyse data from environmental and biometric monitoring devices to predict and autonomously adjust parameters like temperature, lighting, and air quality. Such adaptive systems hold the promise of revolutionising healing landscapes by creating responsive environments that continuously optimise for user health and comfort, ensuring sustained therapeutic benefits in diverse settings [7].

This paper proposes a sophisticated digital twin model enhanced by AI that integrates real-time environmental data and user-specific physiological signals. The framework is designed to replicate the physical attributes of both indoor and outdoor healing landscapes, thereby providing a dynamic, data-driven basis for optimizing environmental conditions. By incorporating advanced machine learning algorithms, the system can predict changes in user comfort and health outcomes, initiating proactive interventions that range from adjusting temperature and ventilation to modifying lighting and even suggesting enhancements in green space design. This approach represents a paradigm shift from static, one-time design interventions to continuously adaptive environments catering to urban populations' nuanced needs.

2. State of the art Review
2.1 Healing Landscapes

Healing environments, commonly known as therapeutic landscapes, are purposefully designed spaces that integrate natural elements, ergonomic principles, and advanced technologies to enhance health outcomes [8]. These environments are increasingly prevalent in healthcare facilities, workplaces, and urban settings, aiming to support physical and psychological well-being. The design of such spaces is grounded in evidence-based practices, emphasising the critical role of environmental factors in promoting recovery, health, and overall wellness [2], [9]. Research consistently underscores the value of well-designed environments in healthcare settings. Access to natural light, green spaces, and tranquil areas has been shown to reduce patient stress and accelerate recovery. Ulrich's study demonstrated that patients with views of nature from hospital windows experienced shorter recovery times and required less pain medication than those with views of urban settings [2]. Similarly, the therapeutic benefits of hospital gardens is highlighted, noting their role in reducing stress and improving patient satisfaction [9]. Studies on indoor environmental quality (IEQ) further emphasise the impact of air quality, acoustics, and aesthetic design on patient outcomes. For instance, a study by Frontczak and Wargocki (2011) found that improved air quality and acoustic comfort in healthcare facilities significantly enhanced patient recovery rates and staff performance [10].

In workplace settings, biophilic design, incorporating natural elements such as plants, water features, and natural light, has been shown to enhance employee satisfaction, cognitive function, and productivity [11]. A study by Nieuwenhuis et al. demonstrated that the presence of indoor plants in offices led to a 15% increase in reported employee well-being and productivity [4]. Additionally, attention to thermal and acoustic comfort is critical for creating supportive work environments [12]. Research by Heerwagen indicates that workplaces designed with biophilic principles and optimised environmental conditions, such as controlled lighting and noise levels, result in reduced stress and improved focus among employees [13]. These findings highlight the measurable benefits of integrating natural and ergonomic elements into workplace design.

Despite the well-documented benefits of healing landscapes, several gaps persist in their design and implementation [14]. Current healing environments are often static, lacking the flexibility to adapt to changing user needs or environmental conditions [15]. For example, fixed lighting and HVAC systems do not account for real-time variations in temperature, noise, or occupancy. While passive technologies are commonly used, active monitoring and dynamic response systems are absent. Real-time adaptability, where lighting, sound, or climate adjust automatically based on sensor data or user feedback, remains underutilised [16].

2.2 Digital Twin Technology in Healthcare and Environmental Control

Digital twin technology, originally developed for aerospace and manufacturing, has become a transformative tool in healthcare and environmental control. A digital twin is a dynamic virtual replica of a physical entity or system, enabling real-time monitoring, analysis, and optimization through integration with advanced technologies such as the IoT, artificial intelligence (AI), edge computing, and big data analytics [6]. In healthcare, digital twins facilitate personalised medicine, surgical planning, and chronic disease management. In environmental control, they optimise building operations, energy efficiency, and occupant well-being. This section explores the applications, benefits, and challenges of digital twin technology in these domains, highlighting its potential to revolutionise health and environmental management.

In healthcare, digital twins have revolutionised personalised medicine and predictive care by creating virtual models of patients that integrate data from wearable devices, electronic health records (EHRS), imaging technologies, and medical sensors. These models provide a comprehensive, real-time view of patient health, enabling proactive interventions. For instance, digital twins can monitor disease progression by analyzing continuous data streams from wearable devices, allowing healthcare providers to predict and prevent adverse health events. A study by El Saddik et al. demonstrated the use of digital twins in chronic disease management, showing how AI-driven predictive models could anticipate diabetic complications, reducing hospital readmissions by 20% [17]. Similarly, digital twins support surgical planning by simulating procedures based on patient-specific anatomical data, improving precision and outcomes [18]. The ability to anticipate health deterioration before it occurs positions digital twins as a cornerstone of preventive healthcare, enhancing patient outcomes and reducing healthcare costs.

In environmental control, digital twins are pivotal in optimizing building management systems (BMS), improving energy efficiency, and enhancing occupant comfort and well-being. By integrating with IoT sensors, AI, and edge computing, digital twins enable continuous monitoring and real-time control of environmental factors such as temperature, humidity, lighting, and air quality[19]. Digital twins play a critical role in maintaining indoor air quality (IAQ) and supporting occupant health. Poor IAQ is linked to respiratory issues, allergies, and reduced cognitive performance [20]. Digital twins address this by continuously monitoring air quality metrics, such as $CO_2$ levels and particulate matter, and adjusting ventilation systems in real time to mitigate health risks. A case study highlighted the use of digital twins in a commercial building, where real-time IAQ monitoring and adaptive ventilation improved occupant satisfaction by 25% and reduced sick building syndrome complaints [7]. Additionally, digital twins can personalise environmental settings based on occupant preferences, such as adjusting lighting or temperature, thereby enhancing comfort and productivity [5].

Furthermore, the integration of AI into health monitoring systems represents a transformative shift in healthcare, enabling advancements in personalised care, predictive analytics, and operational efficiency [21]. AI's predictive capabilities are particularly valuable in personalized interventions [22]. For example, AI algorithms can adjust insulin dosages in real time for diabetic patients based on continuous glucose monitoring, improving glycemic control [23]. By analyzing longitudinal data from wearables, AI can tailor treatment plans to individual patients [24].

Despite its promise, data quality and interoperability remain significant hurdles, as heterogeneous data sources often lack standardised formats, complicating integration [25]. Privacy and security concerns are also critical, as continuous monitoring generates sensitive health data that must be protected against breaches [26]. Additionally, the computational complexity of real-time AI analytics requires robust infrastructure, which may be cost-prohibitive for some healthcare systems.

The combination of AI with digital twins, could further enhance personalized interventions by creating dynamic, patient-specific models that evolve with real-time data [18], [27]. Moreover, advances in edge computing could enable faster, on-device processing of biomedical data, reducing latency and improving accessibility in resource-limited settings [28].

3. Conceptual Framework for AI-Driven Digital Twin Framework for Healing Environments

This dynamic digital twin system is designed to continuously monitor an individual's health status and respond to environmental changes, enhancing overall well-being outcomes.

3.1 Detailed System Architecture

The proposed framework introduces a multi-layered architecture that integrates physical sensing, edge computing, digital twin intelligence, cloud analytics, and user interaction to support predictive health monitoring and intelligent response. There are four layers: physical layer, user interface layer, cloud platform layer, and digital twin layer as shown in Figure 1.

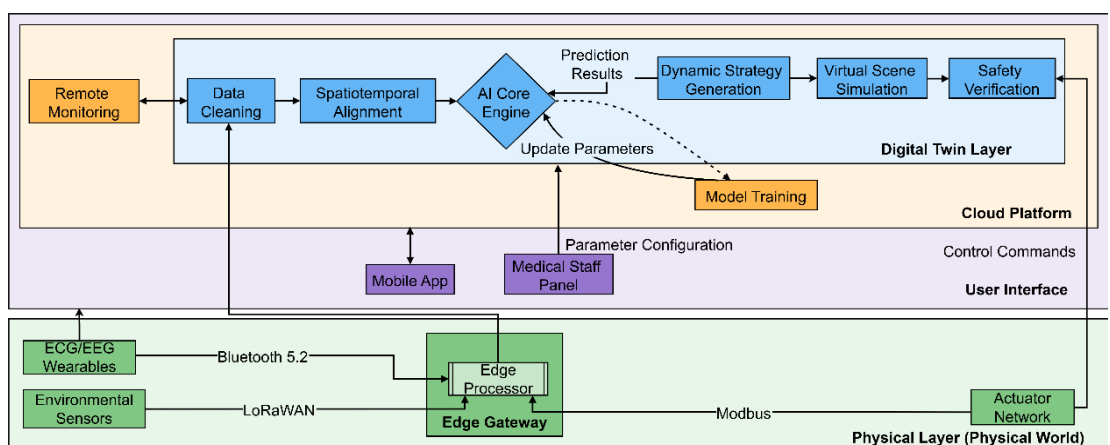

Figure 1 System Architecture

At the physical layer, a range of hardware devices are deployed to capture physiological and environmental data in real time. These include ECG/EEG wearables and environmental sensors, which transmit data using Bluetooth 5.2 and LoRaWAN protocols, respectively. The data is received and aggregated by an edge processor that also connects with the actuator network through Modbus for control signal delivery. This local computation node ensures low-latency communication and pre-filtering before data reaches higher-level processing layers.

The digital twin layer serves as the system's cognitive core. Incoming data is first processed through a pipeline consisting of data cleaning and spatiotemporal alignment modules to ensure quality and coherence. The aligned data is then fed into the AI Core Engine, which performs predictive analysis, such as estimating stress levels or environmental risks. Based on the prediction results, the system generates adaptive control strategies, simulates them in virtual environments, and verifies their safety before issuing control commands back to the physical actuators. This closed-loop process ensures intelligent, safe, and timely intervention in response to real-world conditions.

Complementing the twin layer is the cloud platform, which supports long-term learning and remote visibility. A dedicated module for model training continually updates the AI engine using historical and incoming data, enhancing its predictive performance. Meanwhile, the remote monitoring module provides real-time system oversight, supporting diagnostics and operational transparency.

The user interface layer enables human-system interaction. A medical staff panel allows practitioners to configure system parameters, while a mobile App facilitates remote access for users and caregivers. These interfaces ensure that human decision-makers remain informed and empowered to guide or override automated operations where necessary.

Together, these layers form an integrated cyber-physical-social system capable of autonomously sensing, interpreting, and responding to complex health-related contexts. The architecture supports modular deployment, interoperability through standardized protocols, and continuous learning through AI-driven feedback loops.

3.2 AI-based health monitoring and control system

To bridge the system-level diagram with the AI method flowchart, we highlight the logical zoom-in from the "AI Core Engine" node in the system diagram into the detailed internal mechanisms of the AI model. As shown in Figure 1, the AI Core Engine serves as the centre of prediction and control decisions. Figure 2 expands upon this module, detailing the underlying model architecture and stress classification workflow. The proposed AI framework comprises three key layers which are feature engineering, prediction model, and decision optimization, each responsible for progressively transforming raw input signals into actionable stress-level responses.

To operationalize real-time stress-responsive environment control, a cyclic interaction sequence was implemented across system components. The workflow begins with the acquisition of multimodal physiological and environmental data, which are immediately filtered and feature-engineered at the edge gateway. These features feed into an AI inference module that predicts the current stress level, enabling the generation of structured control commands tailored to the occupant's state. Commands are delivered to actuators, which in turn adjust indoor conditions such as lighting, temperature, sound, and air quality. A feedback mechanism captures actuator states and environmental responses, allowing the AI Core to refine its prediction parameters and control strategies at runtime. This closed-loop interaction enables both reactive and adaptive behaviors, supporting personalized and resilient indoor healing experiences.

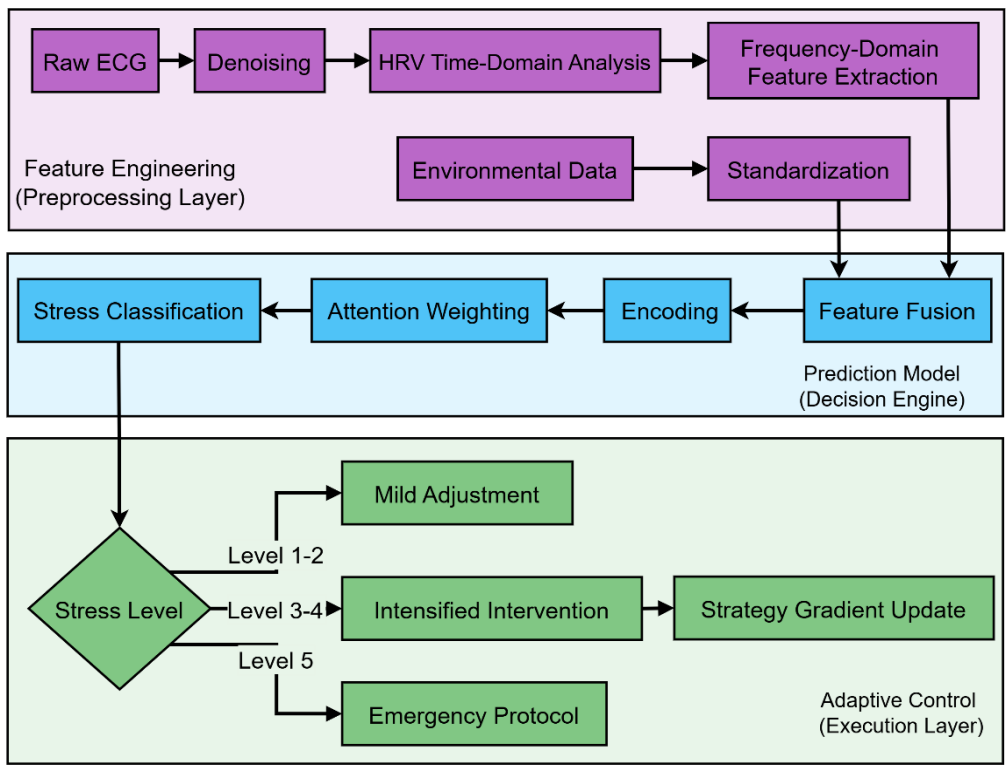

Figure 2 Internal Structure of the AI Core Engine for Stress Prediction and Control Decision

3.3 AI Capabilities for Data-Driven Adaptive Interventions

In the heart of the digital twin system lies a holistic AI model framework that integrates both physiological data and environmental factors to predict health outcomes and influence well-being. An AI-driven digital twin framework in a healing landscape context serves as a dynamic platform that continuously links user-related data (biometrics, health status) and environmental parameters (air quality, light, sound, etc.) to a virtual model of the users' living or working space.

Figure 3 shows the component-level interaction sequence of the AI-based monitoring and control framework. This figure illustrates the real-time interaction between core system components—including sensors, edge processing, AI inference, actuators, cloud services, and the user interface. Dashed arrows represent asynchronous communication or feedback loops, while solid arrows depict real-time, sequential message flow. The system initiates with the sensor transmitting raw ECG and environmental data to the edge processor. The edge device performs preprocessing and forwards structured features to the AI core, which applies machine learning mechanism for continuous risk prediction or classification. Based on the output, a control command is generated and sent to the actuator in JSON format. The actuator returns a feedback signal to the sensor layer. A strategy optimization loop within the AI core is executed every 200 ms to refine decision policies in real-time.

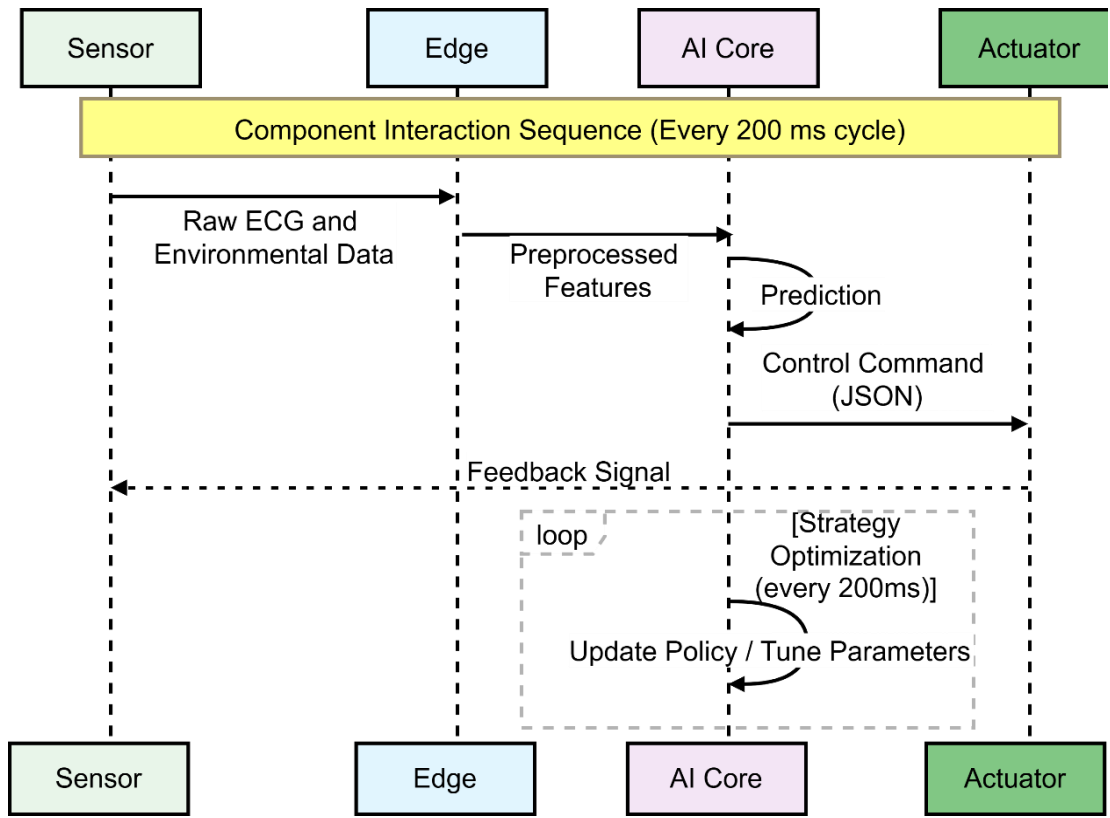

Figure 3 Component Interaction Sequence Diagram

4. Pilot Study

4.1 Objective and Scope

The overarching goal of this pilot study is to critically evaluate and demonstrate the analytical capability of the AI core component within the proposed digital twin framework for health-responsive environmental control. While the broader system envisions real-time interaction between bio-signal monitoring, environmental sensing, and adaptive intervention, the current scope focuses on the stress prediction pipeline due to the absence of dynamic environmental data. To simulate physiological responses under varying environmental influences, the study utilises the open-source MIT-BIH noise stress test database. This dataset provides ECG recordings under different noise conditions, serving as a proxy for real-world stressors.

4.2 Feature Design and Stress Labelling Strategy

The MIT-BIH dataset provides a robust foundation for simulating how noise influences human physiological signals, particularly focusing on cardiac signals as measured by ECG. The dataset contains extensive ECG recordings that have been intentionally corrupted with a range of noise typical of that encountered in ambulatory settings. This includes muscle artefacts, electrode motion artefacts, and other environmental noises. Records 118 and the following 118e6 to 118e24 records are selected.

A clean ECG signal was first filtered using a 0.5–45 Hz bandpass filter to eliminate baseline drift and high-frequency noise. A robust QRS detection algorithm—augmented with parameters for peak prominence and width—was used to identify R-peaks even in challenging signal environments. From these, key HRV metrics were extracted, including standard deviation of NN intervals (SDNN), beats per minute (BPM), corrected QT interval (QTc), and low - to high-frequency spectral power ratio via Welch's method. To handle real-world variability, signals were standardised using z-score normalisation, and RR intervals were filtered based on adaptive bounds to eliminate outliers or missed beats.

For each noisy ECG sample, the same features were computed and compared to the subject's clean baseline, resulting in four relative deviation metrics:

$$\text{rel}_x = \frac{x_{noisy} - x_{baseline}}{x_{baseline} + \varepsilon} \quad (1)$$

In parallel, noise segments were analyzed using both statistical and spectral methods, generating additional descriptors such as:

- Mean, standard deviation, skewness, kurtosis
- Spectral LF/HF ratio of the noise signal

These were used to contextualize environmental factors contributing to stress induction.

A composite stress score was then calculated for each sample as a weighted sum of relative SDNN, BPM, and QTc deviations:

$$\text{Stress Score} = 0.5 \cdot \text{rel}_S\text{DNN} + 0.3 \cdot \text{rel}_B\text{PM} + 0.2 \cdot \text{rel}_Q\text{Tc} \quad (2)$$

A five-level stress classification scheme was adopted based on established HRV references to convert physiological measurements into clinically meaningful stress labels. The classification thresholds were derived from clinical literature on autonomic nervous system balance and dysregulation. As shown in Table 1, each stress level corresponds to specific ranges of SDNN, BPM, QTc, and LF/HF ratio. These stress levels served as supervised learning classification labels and interpretability anchors for downstream model evaluation.

Table 1 Stress level with feature threshold

| Level | Label | SDNN (ms) | BPM | QTc (ms) | LF/HF | State |
|---|---|---|---|---|---|---|
| 1 | Normal | > 50 | 60–80 | < 420 | < 1.5 | Autonomic balance |
| 2 | Mild Stress | 40–50 | 80–90 | 420–440 | 1.5–2.5 | Mild sympathetic activation |
| 3 | Moderate Stress | 30–40 | 90–100 | 440–460 | 2.5–4 | Significant sympathetic tone |
| 4 | High Stress | 20–30 | 100–110 | 460–480 | 4–6 | Parasympathetic suppression |
| 5 | Extreme Stress | < 20 | > 110 | > 480 | > 6 | Autonomic dysfunction |

4.3 Model Training and Evaluation

A random forest classifier was selected for its robustness to feature variance and compatibility with explainability techniques. The dataset was split into training and testing subsets using a 70:30 ratio. To ensure the model's ability to track evolving physiological states, a dynamic sliding-window analysis was implemented. ECG signals were segmented using a 10-second window with a 5-second stride, and features were extracted for each window independently.

Regarding feature reliability, the pipeline adopted a robust feature extraction method that includes enhanced R-peak detection, smoothing, and statistical filtering. Peak detection was designed to adapt to noisy conditions using dynamic thresholds and multi-pass filtering, improving QRS complex identification even in degraded signals.

4.4 Results and Analysis

Figure 4 presents a series of dynamic time-resolved prediction visualisations across six ECG recordings with varying noise conditions. The top subplot shows the temporal variation of heart rate (BPM), while the middle subplot presents the SDNN. The red line shows the BPM variations over time. The BPM signal exhibits clear fluctuations, with periods of relative stability interspersed with sharp transient increases, suggesting acute physiological responses potentially triggered by noise interference or stress events.

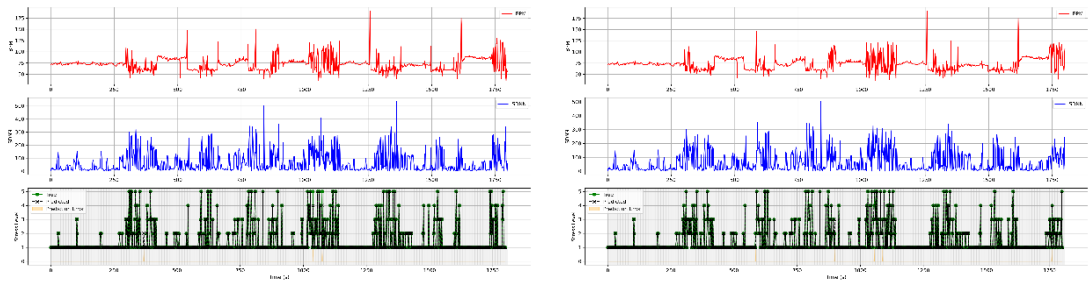

(1) Dynamic prediction for SNR -6    (2) Dynamic prediction for SNR 0

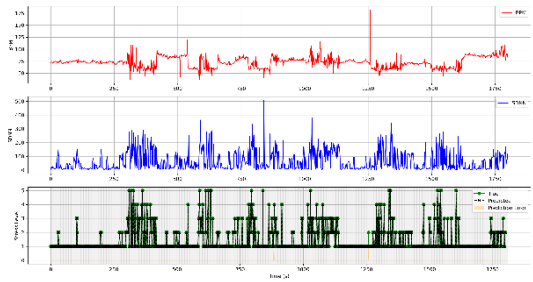
(3) Dynamic prediction for SNR 6

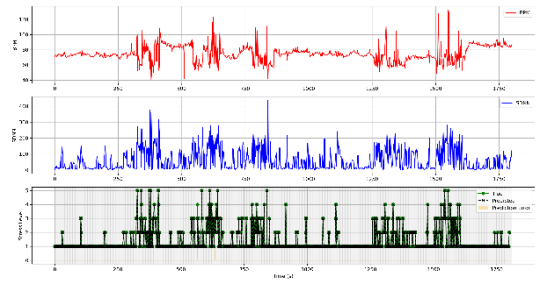
(4) Dynamic prediction for SNR 12

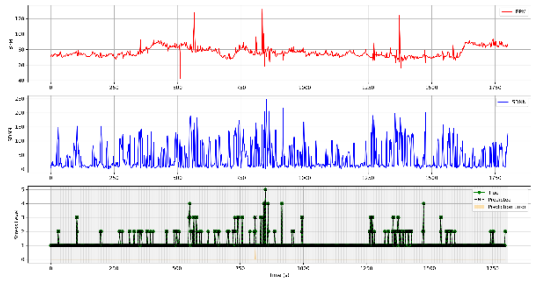
(5) Dynamic prediction for SNR 18

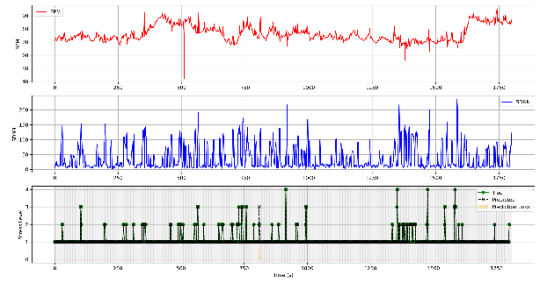
(6) Dynamic prediction for SNR 24

Figure 4 Dynamic prediction for six noise-ECG recordings

The bottom subplot visualises the predicted stress levels compared to the expert rule-based (true) labels, marked by line plots with distinct markers. This subplot compares the true stress levels (green solid line with dots) with the machine learning model predictions (black dashed line with crosses). Minor prediction errors are highlighted by orange shaded regions where absolute differences occur. The BPM and SDNN trajectories provide rich temporal information correlating with stress transitions. Periods of increased BPM combined with reduced SDNN typically coincide with higher predicted stress levels. The dynamic prediction plot demonstrates the model's robustness in tracking real-time physiological stress responses, with very few mismatches between ground truth and prediction. The prediction error visualisation shows that most errors are small and sparse, validating the reliability of the feature-based stress prediction pipeline.

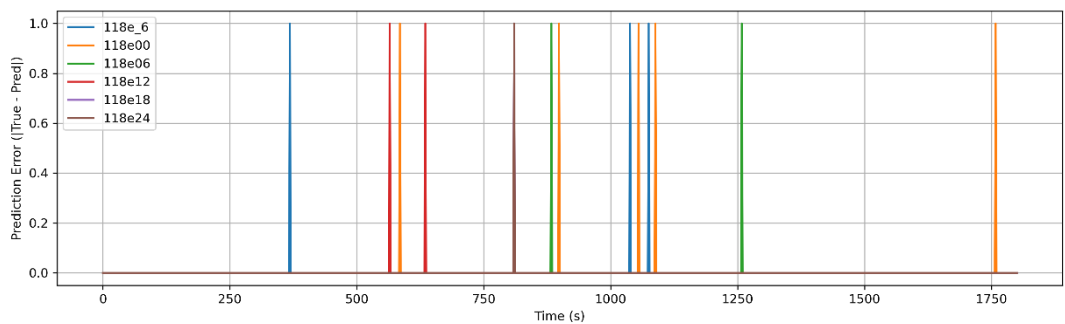
Figure 5 Prediction Error Over Time

Figure 5 illustrates the temporal distribution of prediction errors across all six noisy ECG recordings. Each colour-coded line corresponds to one record, and the vertical spikes represent time windows where the predicted stress level deviates from the reference level derived from expert-defined rules. Despite the inherent challenge of working with noise-stressed physiological data, the model demonstrates overall stability, with most time windows showing zero error (i.e., accurate classification). However, intermittent spikes are observed, most notably in records 118e06, 118e00, and 118e06, suggesting transient mismatches between the model's prediction and the reference classification. These errors are temporally localised and sparse, indicating that the model performs robustly in most signal segments. Records 118e18 and 118e24, for instance, show almost no deviation, highlighting strong predictive consistency under certain noise profiles or intensity conditions. On the other hand, 118e06 exhibits more frequent spikes, indicating that this particular recording posed greater difficulty, possibly due to higher signal distortion or borderline HRV feature values.

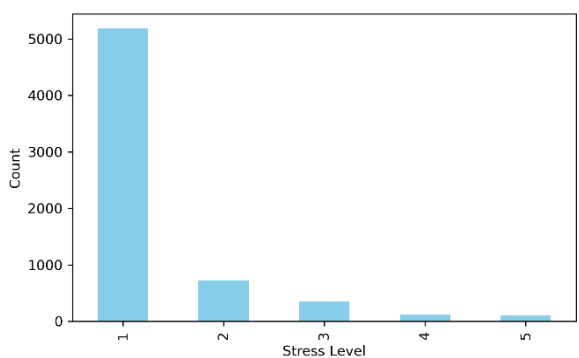

Figure 6 Predicted Stress Level Distribution

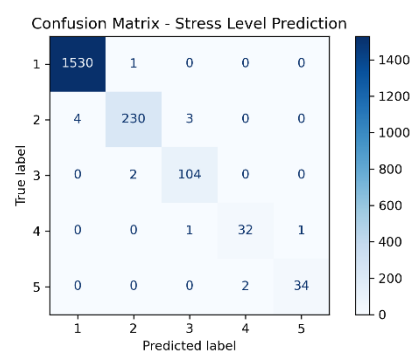

Figure 7 Confusion Matrix for Stress Level Prediction

Figure 6 illustrates the frequency distribution of the predicted stress levels across all classified ECG segments. The histogram reveals a distinct class imbalance in the prediction outcomes, with most instances (over 5,000) being classified as stress level 1, corresponding to a normal or low-stress state. The remaining stress levels, ranging from mild (2) to extreme (5), occur far less frequently, indicating a skewed distribution of classification outcomes.

This imbalance is partially reflective of the actual label distribution, as low-stress segments likely dominate the dataset. However, it also suggests a potential bias in the classifier toward conservative predictions, particularly in ambiguous or noisy signal windows. This phenomenon is common in physiological stress detection where subtle HRV changes might not be sufficient to confidently trigger higher stress class decisions—especially when noise-induced uncertainty is present.

Figure 7 presents the confusion matrix summarizing the performance of the machine learning classifier across five stress levels (1: Normal, 5: Extreme Stress). The diagonal dominance of the matrix indicates strong agreement between predicted and true labels, with the classifier achieving high precision across most categories. Most notably, stress level 1 (normal state) was predicted with exceptional accuracy (1530 correctly classified samples), demonstrating the model's robustness in identifying baseline physiological conditions. For elevated stress levels (2–5), although the sample size is smaller, the model still correctly classifies a large portion of instances. Level 5 shows a small but clean diagonal (34 correct predictions), suggesting the model retains discriminative power even in rare, high-stress states.

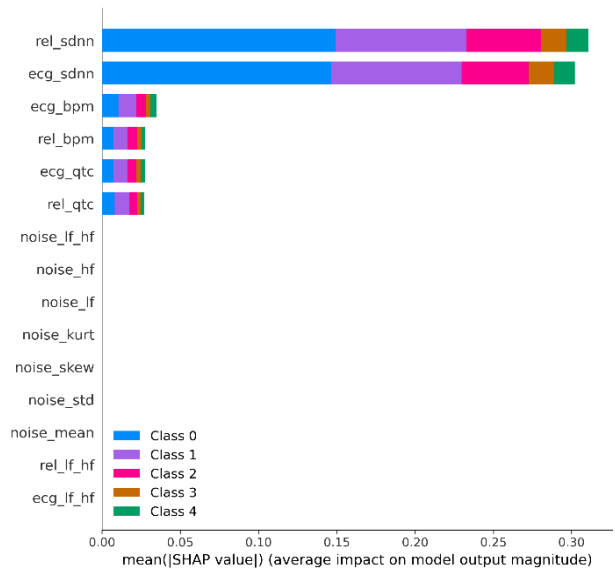

Figure 8 SHAP Feature Importance

Further SHAP value analysis is conducted to provide model transparency (Figure 8 and Figure 9). Figure 8 shows the SHAP summary bar plot, which quantifies the average contribution of each input feature to the model's output across different predicted stress levels. Notably, the relative SDNN (rel_sdnn) and absolute SDNN (ecg_sdnn) emerged as the two most influential features for stress classification across all five classes (from normal to extreme stress), reaffirming the physiological relevance of HRV in stress detection. Other features, such as ECG-derived BPM (ecg_bpm) and QTc interval (ecg_qtc), also contributed meaningfully but to a lesser extent, followed by noise characteristics (e.g., noise_mean, noise_lf, noise_kurt). Interestingly, these noise-derived features were given low SHAP values, indicating that the model predominantly focused on HRV-related features rather than being confounded by ambient noise patterns. This strengthens the evidence that the filtering and feature engineering pipeline successfully isolated meaningful signal dynamics for modelling.

Figure 9 presents the SHAP beeswarm plot for the rel_sdnn feature, which reflects the relative change in SDNN compared to the subject's baseline. This visualization captures both the magnitude and direction of the feature's influence on the predicted stress levels across the dataset. Each dot represents a single sample, with colour encoding the corresponding class (stress level), and position along the x-axis indicating the SHAP value, or the contribution to the model's output.

The plot reveals that those deviations in rel_sdnn, even subtle ones, have a consistent and interpretable effect on the classification outcome. Positive SHAP values (to the right) tend to correspond with higher stress predictions, while negative values (to the left) are associated with lower stress levels. This aligns with physiological expectations: a drop in SDNN is a hallmark of autonomic stress response, particularly reduced parasympathetic activity.

Notably, Class 1 (normal) and Class 2 (mild stress) exhibit tightly clustered SHAP distributions around the origin, suggesting a narrow SDNN variation in low-stress contexts. In contrast, higher stress classes demonstrate greater dispersion, indicative of the model capturing nonlinearity and variability in SDNN deterioration under stress. This non-threshold-based transition enhances the model's sensitivity and robustness, especially in real-world, noisy ECG conditions.

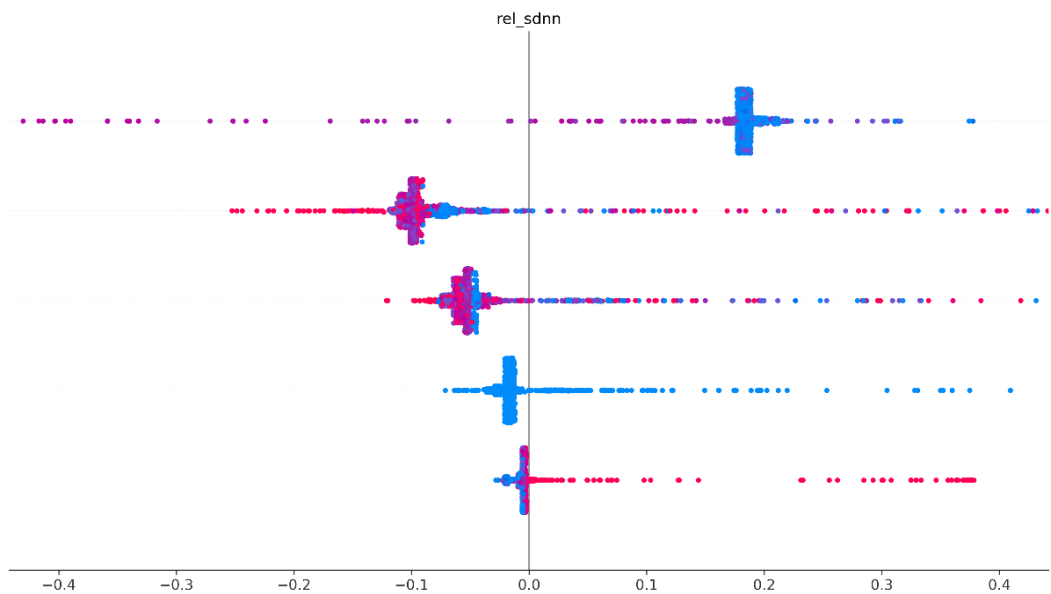

Figure 9 Sensitivity to SDNN

4.5 Environmental Intervention Control

Building upon the model's predictive accuracy, the proposed system translates predicted stress levels into tiered environmental responses via a structured control framework. To this end, a Five-Level Stress–Intervention Mapping was developed (Table 2), which links each discrete stress class directly to context-aware and physiologically grounded environmental control strategies.

This mapping framework reflects a continuum of stress intensity, from homeostatic balance to critical autonomic disruption, and enables the system to respond accordingly. Each predicted class is associated with a specific intervention philosophy: Level 1 (normal) favours stability and minimal intervention, while Level 5 (extreme stress) triggers immersive or emergency containment responses. The corresponding environmental actions span multiple modalities, including lighting, temperature, sound, airflow, aromatherapy, and spatial isolation.

Table 2 Five-Level Stress–Intervention Mapping

| Stress Level | Physiological Indicators (HRV/EEG) | Indoor Interventions | Outdoor Interventions | Theoretical Basis |
|---|---|---|---|---|
| Level 1 (0–0.2) Very Low | HRV > 60 ms EEG: High α waves | • Maintain baseline environment<br>• Mild negative ion release | • Preserve natural ventilation paths<br>• Low-intensity landscape lighting | Homeostatic Maintenance |
| Level 2 (0.2–0.4) Mild | HRV 45–60 ms EEG: Mid α waves | • Dynamic sunrise lighting (3000K)<br>• Background white noise (45 dB) | • Activate shallow water flow<br>• Trigger aromatic plant airflow | Ulrich, 1991 |
| Level 3 (0.4–0.6) Moderate | HRV 30–45 ms EEG: High β waves | • Biowall oxygen-boosting mode<br>• Local temperature gradient (±2°C)<br>• Directional natural soundscape | • Intermittent fog system<br>• Dynamic tree canopy shading | Kaplan, 1995 |
| Level 4 (0.6–0.8) High | HRV 20–30 ms EEG: Low γ waves | • Full-spectrum light therapy<br>• Whole-body vibration (40Hz)<br>• Emergency mist cooling | • Enhanced waterfall mode<br>• Activate magnetic walking trail | Sternberg, 2009 |
| Level 5 (0.8–1.0) Critical | HRV < 20 ms EEG: Abnormal discharge | • Emergency pod activation (isolation + VR natural scene)<br>• Temporary hyperbaric oxygen support | • Open emergency shelter kiosks<br>• Deploy drones with aromatic diffusers | WHO Guidelines |

Importantly, these responses are deployed across a multi-scale architecture encompassing personal, room, building, and landscape levels (Table 3). This allows for immediate relief (e.g., wearable cooling at the personal scale) and ambient recalibration (e.g., fog dispersion or vegetation control at the landscape scale). This design ensures that interventions are both biologically relevant and spatially appropriate, allowing for scalable and coordinated stress recovery mechanisms.

At the personal scale, rapid-response actuators such as wearable cooling bands or biofeedback patches provide immediate physiological relief, often within seconds of detection. The room scale incorporates perceptual modulation techniques such as circadian lighting adjustment, localized soundscapes, or biophilic visual cues (e.g., animated walls or projected nature scenes). These aim to modulate the sensory environment without disrupting spatial continuity. At the building scale, zone-specific HVAC modulation and smart façade systems offer broader thermal and air quality control. Meanwhile, the landscape scale addresses collective stress trends in open environments using mist systems, dynamic vegetation shading, water features, or aroma-delivering drones. These interventions operate with longer latencies but higher spatial reach, enabling group-level restorative effects in shared spaces. The system achieves temporal-spatial alignment between stress detection and environmental healing by orchestrating responses across these spatial layers.

Table 3 Multi-Scale Intervention Strategy

| Scale | Intervention Mode | Response Time |
|---|---|---|
| Personal Scale | Wearable microclimate modulation (e.g., cooling neckband) | < 5 s |
| Room Scale | Localized light and sound field reconstruction | 15–30 s |
| Building Scale | Global HVAC mode switching | 1–5 min |
| Landscape Scale | Coordinated water and vegetation control | 5–15 min |

5. Conclusion

This research proposes an AI-enhanced digital twin framework that redefines the role of built environments in supporting mental and physiological well-being. The system enables active regulation of environmental conditions based on biosignal-derived insight by embedding real-time stress detection and adaptive control mechanisms into a digital representation of physical space. Rather than functioning as a passive replica, the digital twin becomes a cyber-physical mediator of health, capable of dynamically modulating lighting, sound, airflow, and sensory stimuli to align with the occupant's evolving stress state.

The system's core is a machine learning pipeline trained on ECG-derived physiological features, particularly SDNN, BPM, and QTc, extracted from the MIT-BIH Noise Stress Test dataset. SHAP-based explainability further enhances the model's transparency, confirming the physiological validity of its stress classifications. These predictions are operationalised through a clinically grounded Five-Level Stress–Intervention Mapping, which connects model output to a structured set of spatial interventions. A Multi-Scale Intervention Strategy ensures that control logic is distributed across personal, room, building, and landscape layers, allowing interventions to be targeted and scalable. The case study validates that subtle biosignal shifts can reliably trigger intelligent environmental responses.

Beyond this technical implementation, the framework represents a significant step toward user-centred and health-responsive design. By marrying biophilic design principles with AI analytics, natural environmental affordances, e.g., daylight, green space, fresh air, and acoustics, are transformed from qualitative comforts into quantifiable, controllable variables. In healing environments such as hospitals, these features can be continuously optimised in response to patient-specific physiological states, enabling therapeutic spaces that adapt in real time to enhance comfort, recovery, and psychological resilience.

Looking forward, this work opens new directions for research and deployment. Future studies will explore multimodal biometric integration (e.g., EEG, GSR), closed-loop user feedback, and large-scale deployment in smart buildings and healthcare settings. Ultimately, the convergence of digital twins, physiological AI, and environmental design offers a scalable pathway toward intelligent, emotionally responsive, and health-centred architectural systems.